 \documentclass[preprint2]{aastex}

\def\dsm{$M_\odot$}
\def\dsr{$R_\odot$}
\def\dsl{$L_\odot$}
\def\teff{$T_{\rm eff}$}
\def\dov{$\delta_{\rm ov}$}
\def\dra{$r_{01}$}
\def\drb{$r_{10}$}

\def\dhz{$\mu$Hz}

\def\dnu0{$\nu_{0}$}
\def\dcc{$\chi_{\mathrm{c}}^{2}$}
\def\dcn{$\chi_{\mathrm{\nu}}^{2}$}

\shorttitle{Estimate the radius of the convective core}
\shortauthors{Wuming Yang}

\begin{document}


\title{Estimate the radius of the convective core of main-sequence
stars from observed oscillation frequencies}

\author{Wuming Yang}
\affil{Department of Astronomy, Beijing Normal University,Beijing 100875, China}
\email{ yangwuming@bnu.edu.cn; yangwuming@ynao.ac.cn}


\begin{abstract}
The determination of the size of the convective core of main-sequence 
stars is usually dependent on the construction of models of stars. 
Here we introduce a method to estimate the radius of the convective core
of main-sequence stars with masses between about 1.1 and 1.5 $M_{\odot}$
from observed frequencies of low-degree p-modes. A formula is proposed
to achieve the estimation. The values of the radius of the convective 
core of four known stars are successfully estimated by the formula.
The radius of the convective core of KIC 9812850 estimated by the 
formula is $\mathbf{0.140\pm0.028}$ $R_{\odot}$. In order to confirm 
this prediction, a grid of evolutionary models were computed.
The value of the convective-core radius of the best-fit model 
of KIC 9812850 is $0.149$ $R_{\odot}$, which is in good 
agreement with that estimated by the formula from observed frequencies.
The formula aids in understanding the interior structure of stars
directly from observed frequencies. The understanding is not 
dependent on the construction of models.
\end{abstract}

\keywords{Convection --- stars: evolution --- stars: interiors --- 
stars: oscillations (including pulsations)}

\section{INTRODUCTION}

By matching the luminosity, atmospheric parameters, and oscillation 
frequencies of models with the observed ones, 
asteroseismology is used to determine fundamental parameters
of stars. Asteroseismology is also used to probe physical processes 
in stars and diagnose internal structures of stars \citep{roxb94,
roxb99, roxb01, roxb03,roxb04, roxb07, cunh07, cunh11, bran10, bran14, 
chri10, dehe10, yang10a, yang12, yang15, silv11, silv13, chap14, 
ge14, guen14, liu14, yang16}. Asteroseismology is a powerful tool 
for studying the structure and evolution of stars.
 
Stars with a mass larger than $1.1$ \dsm{} are considered 
to have a convective core during their main sequence (MS) stage.
Due to the fact that the overshooting of the convective core
can bring more hydrogen-rich material into the core,
the evolution of a star could be significantly affected by
the overshooting. Thus determining the size of the convective
core including the overshooting region is important for
understanding the structure and evolution of stars. However,
the size of the convective core has never been determined
directly from observed data of stars. Generally,
the understanding of the size of the convective core derives
from the computation of evolutionary models of stars.

When seeking to probe the internal structures of 
stars with low-$l$ p-modes, the small separations, $d_{10}$,
$d_{01}$, $d_{02}$, and $d_{13}$, and the ratios of the small 
separations to the large separations, \drb{}, \dra{}, $r_{02}$,
and $r_{13}$ \citep[and references therein]{roxb03, yang07b}, 
are considered to be the very useful diagnostic tools. The small 
separations $d_{10}$ and $d_{01}$ are defined as \citep{roxb03}
\begin{equation}
d_{10}(n)=-\frac{1}{2}(-\nu_{n,0}+2\nu_{n,1}-\nu_{n+1,0})
\label{d10}
\end{equation}
and
\begin{equation}
d_{01}(n)=\frac{1}{2}(-\nu_{n,1}+2\nu_{n,0}-\nu_{n-1,1}).
\label{d01}
\end{equation}
But in calculation, the smoother five-point separations 
are adopted. 

Stars with masses between about 1.1 and 1.5 \dsm{} have
a convective core during their MS. The discontinuity
in density at the edge of the convective core increases with 
the evolution of the stars. The rapid variation of density 
with depth in a stellar core can distort acoustic wave 
propagation in stellar interiors, producing a reflected
wave \citep{roxb07}. The reflectivity can come from the rapid 
density change at the edge of the convective core \citep{roxb07}. 
For the modes with frequencies larger than a critical frequency,
they can penetrate into the convective core. 
Partial wave reflection at the core boundary
could lead to acoustic resonances in the convective core \citep{roxb04}.
As a consequence, at high frequencies, we would see a periodic
variation in the small separations with frequency \citep{roxb04}. 
If this periodic component is determined from
observations, it can be used for constraining the size of 
the convective core \citep{roxb99, roxb01, roxb04}. 

\cite{roxb94, roxb00a, roxb00b, roxb01} developed the theory of 
semiclassical analysis that can more accurately describe the low-degree 
p-modes and the small separations. However, \cite{roxb04} pointed out
that their expressions for the perturbations in the phase shifts 
are not transparent enough to serve as a basis for simple estimates. 
The effects of the convective core on $d_{10}$, $d_{01}$,
\drb{}, and \dra{} are also studied by other authors 
\citep{cunh07, cunh11, bran10, bran14, dehe10, silv11, silv13, liu14, yang15}.
The conclusion is that the ratios \dra{} and \drb{} 
can be affected by the presence of the convective core.

In order to isolate the frequency perturbation produced 
by the edge of the convective core, \cite{cunh07} and \cite{cunh11}
defined a tool $r_{0213}=r_{02}-r_{13}$. They have shown 
that the tool can potentially be used to infer information about 
the amplitude of the discontinuity in the sound speed at the edge 
of the convective core but it is unable to fully isolate the frequency
perturbation. 

\cite{yang15} show that the ratios \dra{} and \drb{} of a star 
with a convective core can be described by equation
\begin{equation}
B(\nu_{n,1})=\frac{2A\nu_{n,1}}{2\pi^{2}(\nu^{2}_{0}-\nu^{2}_{n,1})}
\sin(2\pi\frac{\nu_{n,1}}{\nu_{0}})+B_{0},
\label{eqb}
\end{equation}
where the quantities $A$ and $B_{0}$ are two parameters, the $\nu_{0}$
is the frequency of $l=1$ mode whose inner turning point
is located on the boundary between the radiative
region and the overshooting region of the convective core. 

In this work, we propose a method to estimate the radius of the 
convective core of MS stars with masses between about 1.1
and 1.5 \dsm{} from observed frequencies of low-degree
p-modes. The estimated radius is comparable with that obtained 
from evolutionary model. Individual frequencies of p-modes of
KIC 9812850 have been extracted by \cite{appo12}. 
The mass of KIC 9812850 estimated by \cite{metc14}
is $1.39\pm0.05$ \dsm{}. Thus KIC 9812850 could have a convective
core. We determined the radius of the convective core of KIC 9812850
in two ways. One is estimated from the observed frequencies; 
the other is determined from the best model for KIC 9812850.
In Section 2, a formula that can be used to determine the radius
of the convective core from oscillation frequencies is proposed
and is applied to different stars. In Section 3, based on 
finding the maximum likelihood of models of a grid of evolutionary
tracks, the best-fit model of KIC 9812850 is found out; 
and then, we compare the radius of the convective core of
the best model with that determined from oscillation frequencies.
Finally, we give a discussion about the domain of the validity
of the method and summary in Section 4. 

\section{ESTIMATE THE RADIUS OF THE CONVECTIVE CORE FROM OSCILLATION FREQUENCIES }

The inner turning point, $r_{t}$, of the mode with a frequency 
$\nu_{n,l}$ is determined by
\begin{equation}
r_{t}=f_{0}\frac{c_{s}(r_{t})}{\nu_{n,l}}\frac{\sqrt{l(l+1)}}{2\mathrm{\pi}},
\label{tp}
\end{equation}
where $c_{s}(r_{t})$ is the adiabatic sound speed at radius $r_{t}$,
the value of the parameter $f_{0}$ is $2.0$ \citep{liu14}. For the modes 
with $l=1$, the frequency $\nu_{0}$ of the mode whose inner turning point
is just located on the boundary between the radiative region and the 
overshooting region of the convective core can be estimated by 
equation (\ref{eqb}) from observed frequencies and the ratios computed 
from the frequencies. Thus the radius of the convective core including 
the overshooting region, $r_{c}$, can be determined by
\begin{equation}
r_{c}=\frac{c_{s}(r_{c})}{\nu_{0}}\frac{\sqrt{2}}{\mathrm{\pi}}.
\label{rc1}
\end{equation}
In the middle stage of MS stars with masses between about 1.1 
and 1.5 \dsm{}, the magnitude of $c_{s}(r_{c})$ is of the order of 
about $5\times10^{7}$ cm s$^{-1}$. Thus for the MS stars 
extracted individual frequencies of low-degree p-modes, combining 
the equation (\ref{eqb}), the radius $r_{c}$ can be estimated by
\begin{equation}
  \begin{array}{ll}
r_{c} & \approx \frac{5\times10^{7}}{\nu_{0}\times10^{-6}} 
        \frac{\sqrt{2}}{\mathrm{\pi}}\frac{1}{6.9598\times10^{10}}\\
      & = \frac{323}{\nu_{0} (\mu \rm{Hz})} (R_{\odot})\\
\end{array}
\label{rc2}
\end{equation}
from the observed data.

\textbf{By using the function of nonlinear curve fitting of Origin software
where chi-square fitting is used and Hessian matrix is calculated
\footnote{http://www.originlab.com/doc/Origin-help/NLFit-theory},}
from the observed frequencies \citep{beno09} and ratio \drb{} of HD 49933, 
\textbf{we obtained} that the value of \textbf{the parameter $\nu_{0}$ 
of equation (\ref{eqb})} is $1920\pm46$ \dhz{} \textbf{for HD 49933}.
From the observed frequencies of KIC 6225718 \citep{tian14}
and ratio \drb{}, the value of $\nu_{0}$ is estimated to be about $5764\pm312$ \dhz{}.
From the frequencies of KIC 2837475 and KIC 11081729 given by \cite{appo12},
the value of $\nu_{0}$ is estimated to be $913\pm36$ \dhz{} for KIC 2837475  
and $795\pm21$ \dhz{} for KIC 11081729. 
 
Using formula (\ref{rc2}), one can obtain that the radius of the convective
core including overshooting region is about $0.056$ \dsr{} for KIC 6225718,
$0.168$ \dsr{} for HD 49933, $0.354$ \dsr{} for KIC 2837475, 
and $0.406$ \dsr{} for KIC 11081729. The models of the
four stars were determined \citep{tian14, liu14, yang15, yang15b}. 
The radius of the convective core including overshooting 
region is around $0.056$ \dsr{} for the model 14 of KIC 6225718 \citep{tian14},
$0.17$ \dsr{} for the model M52 of HD 49933 \citep{liu14}, $0.358$ \dsr{} for 
the model Ma14 of KIC 2837475 \citep{yang15}, and $0.37$ \dsr{} for the model Ma6
of KIC 11081729 \citep{yang15b}. The values of radius $r_{c}$ of these stars,
estimated from observed data, are in good agreement with those obtained from 
the models of the stars. This indicates that the radius of the convective 
core $r_{c}$ of the MS stars could be determined directly from observed
frequencies and ratios \dra{} and \drb{} by using equations (\ref{eqb}) and (\ref{rc2}).

\cite{appo12} extracted the frequencies of low-$l$ p-modes of KIC 9812850. 
Using the ratio \drb{} computed from the frequencies and the observed 
frequencies $\nu_{n,1}$, the values of $A$, $\nu_{0}$, and $B_{0}$
in equation (\ref{eqb}) are estimated to be $264\pm62$ $\mathrm{\pi}$, 
$2309\pm141$ \dhz{}, and $0.043\pm0.005$, respectively, for KIC 9812850.
Thus the radius $r_{c}$ of KIC 9812850 is predicted to be about $0.14$ \dsr{}
by formula (\ref{rc2}).

\section{THE RADIUS OF THE CONVECTIVE CORE OF THE BEST MODEL OF KIC 9812850}
\subsection{Evolutionary Models}
In order to compare the value of $r_{c}$ estimated by equations 
(\ref{eqb}) and (\ref{rc2}) for KIC 9812850 with that of evolutionary model
of KIC 9812850, we sought for the best model of KIC 9812850 that
match both non-seismic constraints and seismic characteristics in
a grid of evolutionary models. We used the Yale Rotation Evolution 
Code (YREC) \citep{pins89, yang07a, yang15} to construct the models. 
For the microphysics, the OPAL equation-of-state table EOS2005 \citep{roge02}
and OPAL opacity table GN93 \citep{igle96} were adopted, supplemented by
the \cite{alex94} opacity tables at low temperature. The models with
a mass less than 1.30 \dsm{} take into account the diffusion and settling
of both helium and heavy elements by using the diffusion coefficients 
of \cite{thou94}. The standard mixing-length theory is adopted to treat
convection. The mixing-length parameter $\alpha$ is a free parameter
in this work. For the Sun, the value of the $\alpha$ for the YREC is $1.74$.
The distance of the overshooting of the convective core is defined as
\dov{}$H_{p}$, where \dov{} is a free parameter and $H_{p}$ is the
local pressure scale-height. The full mixing of material is assumed 
in the overshooting region. The initial helium mass fraction is fixed
at the standard big bang nucleosynthesis value 0.248 \citep{sper07}
and 0.295. All models are evolved from zero-age MS to the end of MS.
The values of the input parameters, mass, $\alpha$, \dov{}, and
$Z_{i}$ for the calculations are summaried in Table \ref{tab1}.

The adiabatic oscillation frequencies $\nu_{n,l}$ of models 
were computed by using the pulsation code jig7 of \cite{guen94}. 
The effects of the near-surface effects of a model on the frequencies
were calculated by using the method of \cite{kjel08}.

\subsection{Observational Constraints on Models }

KIC 9812850 is an F8 star \citep{wright03}. The value of
[Fe/H] given by \cite{ammo06} is $0.00^{+0.15}_{-0.16}$,
but that given by \cite{brun12} is $-0.16\pm0.06$ for KIC 9812850.
Combining the value of $0.023$ of $(Z/X)_{\odot}$ of the Sun, 
the value of $(Z/X)_{\mathrm{s}}$ of KIC 9812850 is estimated to
be between 0.016 and 0.033 for the [Fe/H] of \cite{ammo06},
or in the range of 0.014 and 0.018 for the [Fe/H] of \cite{brun12}.
The effective temperature of KIC 9812850 is $6297\pm70$ K 
\citep{ammo06} or $6330\pm70$ K \citep{brun12}.
The estimated atmospheric parameters of stars hotter than 6,000 K 
could be affected by the method of spectral analysis \citep{mol13}.
Therefore, the atmospheric parameters determined by both \cite{ammo06} 
and \cite{brun12} were considered in this work.

The parallax of KIC 9812850 is in the range between about 5.9 
and 17.5 mas \citep{khar01, ammo06}. The bolometric correction of
KIC 9812850 is estimated from the tables of \cite{flow96}. 
The extinction of KIC 9812850 is given by \cite{ammo06}. The 
visual magnitude of this star is $9.5\pm0.3$ mag \citep{droe06, ammo06}.
Thus the luminosity of KIC 9812850 is estimated to be about $2.6\pm2.3$ \dsl{}.

In order to find the best model for KIC 9812850, we calculated
the likelihood function of all models. The likelihood function is defined
as \citep{basu10}
\begin{equation}
\mathcal{L} = \frac{1}{(2\pi)^{N/2}\prod_{i=1}^{N}\sigma(C_{i}^{ob})}\exp(-\frac{1}{2}\chi^{2}),
\label{likel}
\end{equation}
where
\begin{equation}
\chi^{2} = \sum_{i=1}^{N}[\frac{C_{i}^{\mathrm{th}}-C_{i}^{\mathrm{ob}}}{\sigma(C_{i}^{\mathrm{ob}})}]^{2},
\end{equation}
the quantity $C_{i}^{\mathrm{ob}}$ indicates the observed \teff{},
$L/L_{\odot}$, $(Z/X)_{\mathrm{s}}$, and $\nu_{n,l}$, while the $C_{i}^{\mathrm{th}}$
corresponds to the \teff{}, $L/L_{\odot}$, $(Z/X)_{\mathrm{s}}$, and $\nu_{n,l}$
of models. The quantity $\sigma(C_{i}^{\mathrm{ob}})$ represents the 
observational error of $C_{i}^{\mathrm{ob}}$. The value of $N$ is 45. 

Moreover, the values of classical \dcc{} and \dcn{}
of models were also computed as a reference. The \dcc{} and \dcn{} are defined as
\begin{equation}
\chi^{2}_{\mathrm{c}} = \frac{1}{3}\sum_{i=1}^{3}
[\frac{C_{i}^{\mathrm{th}}-C_{i}^{\mathrm{ob}}}{\sigma(C_{i}^{\mathrm{ob}})}]^{2}
\label{chic}
\end{equation}
and
\begin{equation}
\chi_{\nu}^{2} = \frac{1}{42}\sum_{i=1}^{42}
[\frac{\nu_{i}^{\mathrm{th}}-\nu_{i}^{\mathrm{ob}}}{\sigma(\nu_{i}^{\mathrm{ob}})}]^{2},
\label{chinu}
\end{equation}
respectively, where $C_{i}=(T_{\mathrm{eff}}, L/L_{\odot}, (Z/X)_{\mathrm{s}})$ 
and $\sigma(C_{i}^{\mathrm{ob}})$ denotes the observational error,
$\nu_{i}$ corresponds to frequencies. The observational error
of $\nu_{i}^{\mathrm{ob}}$ is indicated by $\sigma(\nu_{i}^{\mathrm{ob}})$. 

When the model evolves to the vicinity of the error-box of luminosity 
and effective temperature in the Hertzsprung-Russell diagram, 
the time-step of the evolution for each track is set as small 
as $1$ Myr, which ensures that the consecutive models have an
approximately equal \dcn{}.

\subsection{The Best Models of KIC 9812850}
For a given mass, the model that maximizes $\mathcal{L}$
is chosen as a candidate for the best-fit model. Table \ref{tab2} 
lists four models which have a larger $\mathcal{L}$ in 
the calculations and shows that model M3 has the maximum $\mathcal{L}$. 

\begin{figure}
\centering
\includegraphics[scale=0.5, angle=-90]{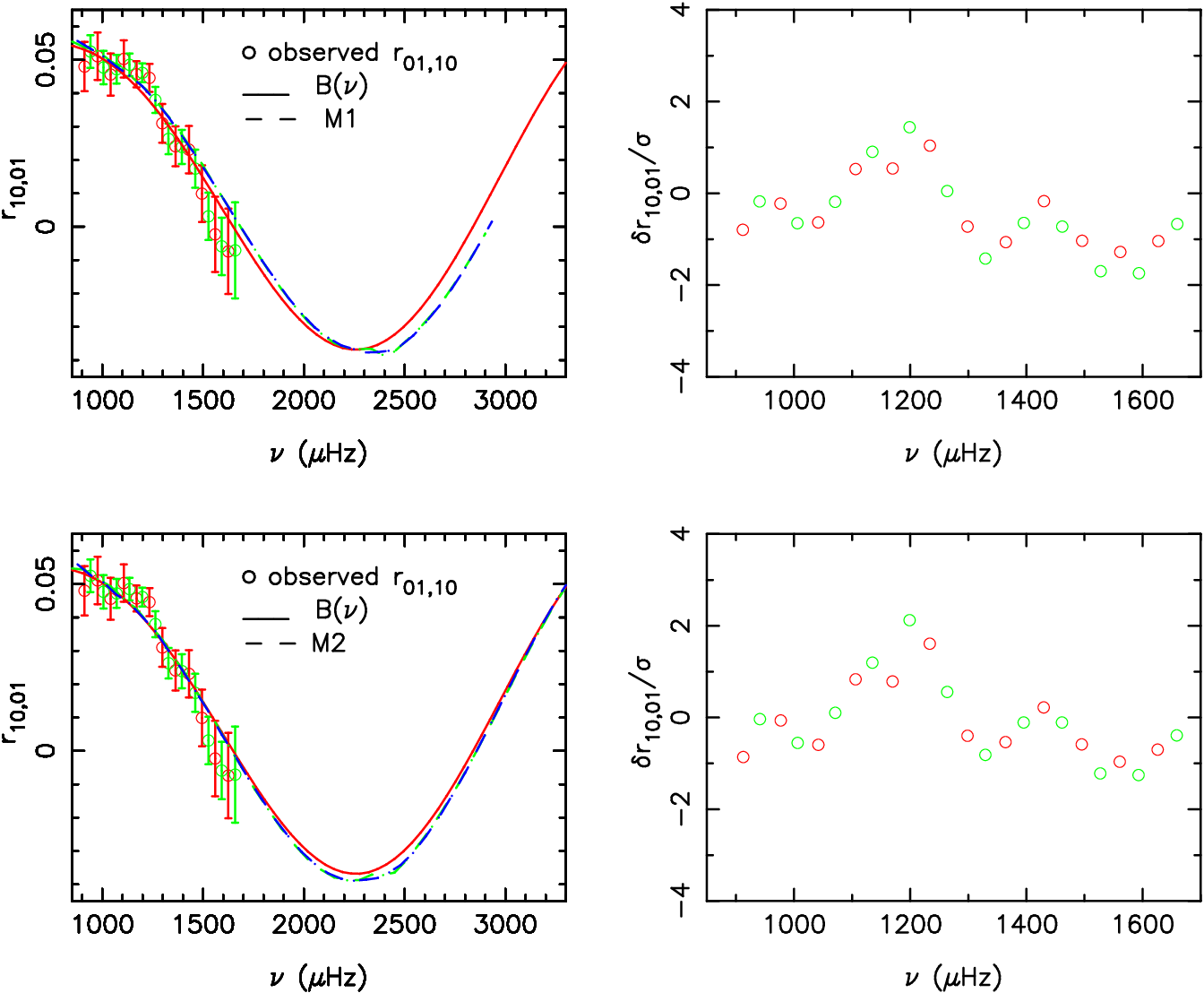}
\includegraphics[scale=0.5, angle=-90]{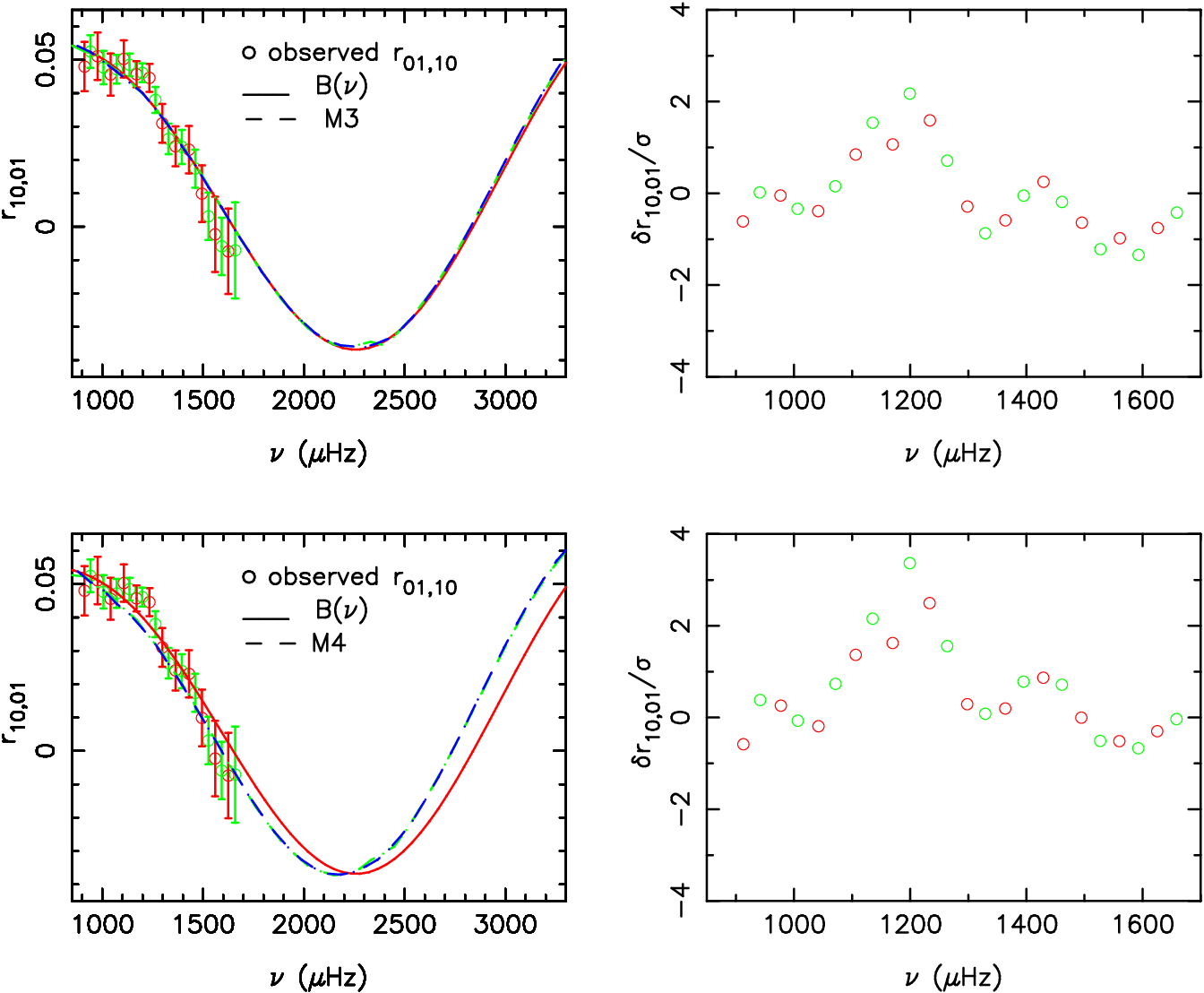}
\caption{Left panels show the distributions of ratios 
\dra{} and \drb{} and the results of equation (\ref{eqb}) as a
function of frequency. The values of $A$, $\nu_{0}$, and $B_{0}$ 
for $B(\nu)$ are $180\pi$, $2180$ \dhz{}, and $0.0465$, respectively.
Right panels represent differences between observed \dra{} and \drb{}
and those of models, in the sence (Observed value $-$ Model value)
/Observational error.}
\label{fig1}
\end{figure}

Figure \ref{fig1} compares the distributions of the observed \dra{} 
and \drb{} with those calculated from the models listed in Table \ref{tab2}. 
The distributions of \dra{} and \drb{} of KIC 9812850 are reproduced 
well by models M2 and M3. The \textbf{right} panels of Figure \ref{fig1} 
show that there are periodic variations in the differences between
the observed ratios and those of models. This may come from the effects
of the helium ionization region and the base of the convective envelope 
on the observed frequencies \citep{roxb04, mazu14}. Model M3 not only 
maximizes the likelihood function in the calculations, but reproduces 
the distributions of observed \dra{} and \drb{} of KIC 9812850. 
Therefore, M3 is chosen as the best-fit model of KIC 9812850. 

Moreover, Figure \ref{fig1} shows that the distributions of \dra{} and \drb{} 
of M3 are reproduced well by equation (\ref{eqb}) with $A=180\pi$, 
$\nu_{0}=2180$ \dhz{}, and $B_{0}=0.0465$. This indicates that ratios \dra{}
and \drb{} can be described by equation (\ref{eqb}). 

The value of $r_{c}$ of model M3 is $0.149$ \dsr{}, which is in good 
agreement with $\mathbf{0.140\pm0.028}$ \dsr{} estimated by equation (\ref{rc2}). 
The radius of the convective core including overshooting region of 
the best model of KIC 9812850 is successfully estimated by equations
(\ref{eqb}) and (\ref{rc2}) from observed oscillation frequencies.

Moreover, the value of $\nu_{max}$ of KIC 9812850 is $1186$ \dhz{} 
\citep{appo12}, which is less than $2309$ \dhz{} of $\nu_{0}$. 
The value of \dov{} of M3 is $0.2$, which is consistent
with the deduction that if the value of $\nu_{max}$ of a star
is less than the value \textbf{of} $\nu_{0}$, the star may have 
a small \dov{} \citep{yang15b}.

\section{DISCUSSION AND SUMMARY}
\subsection{Discussion}
\begin{figure}
\centering
\includegraphics[scale=0.5, angle=-90]{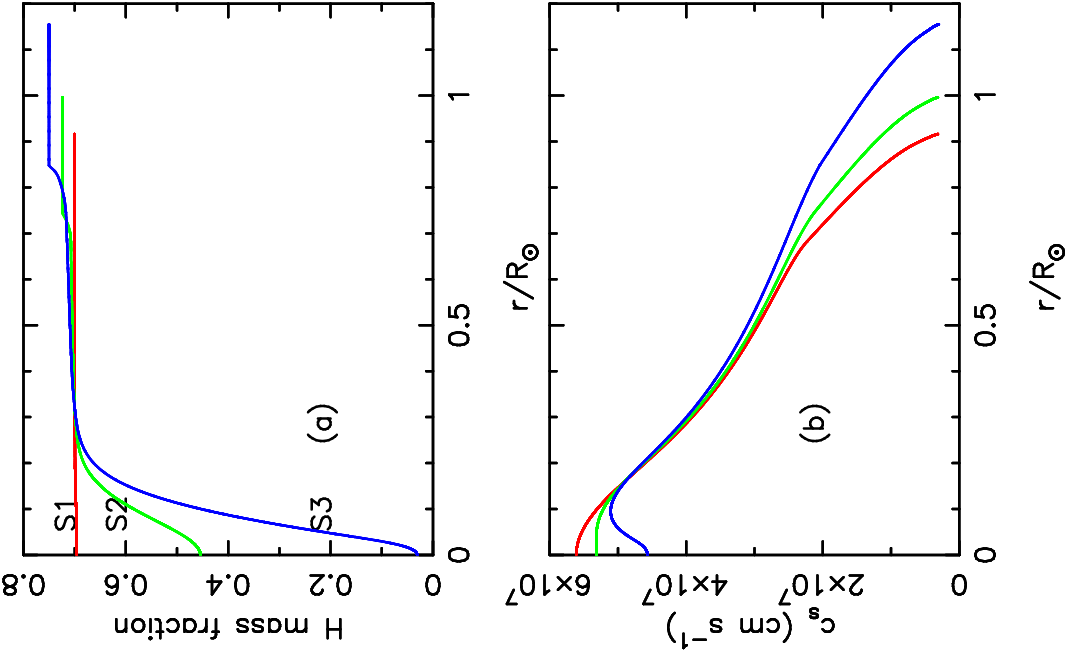}
\includegraphics[scale=0.5, angle=-90]{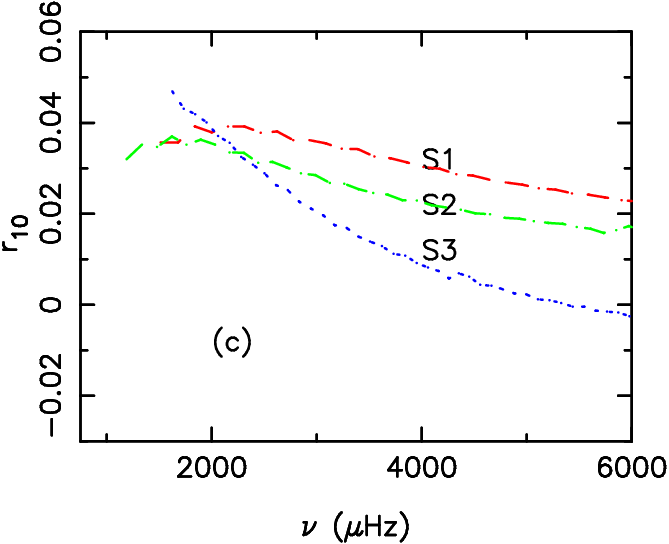}
\caption{Panels (a) and (b) show radial distributions of H mass fraction 
and adiabatic sound speed of models with $M=1.04$ \dsm{}, respectively.
The symbols S1, S2, and S3 indicate different evolutionary stages. \textbf{ 
The cores of the models are radiative. } Panel (c) represents 
distributions of \drb{} of the models as a function of frequency.} 
\label{fig2}
\end{figure}

\begin{figure}
\centering
\includegraphics[scale=0.5, angle=-90]{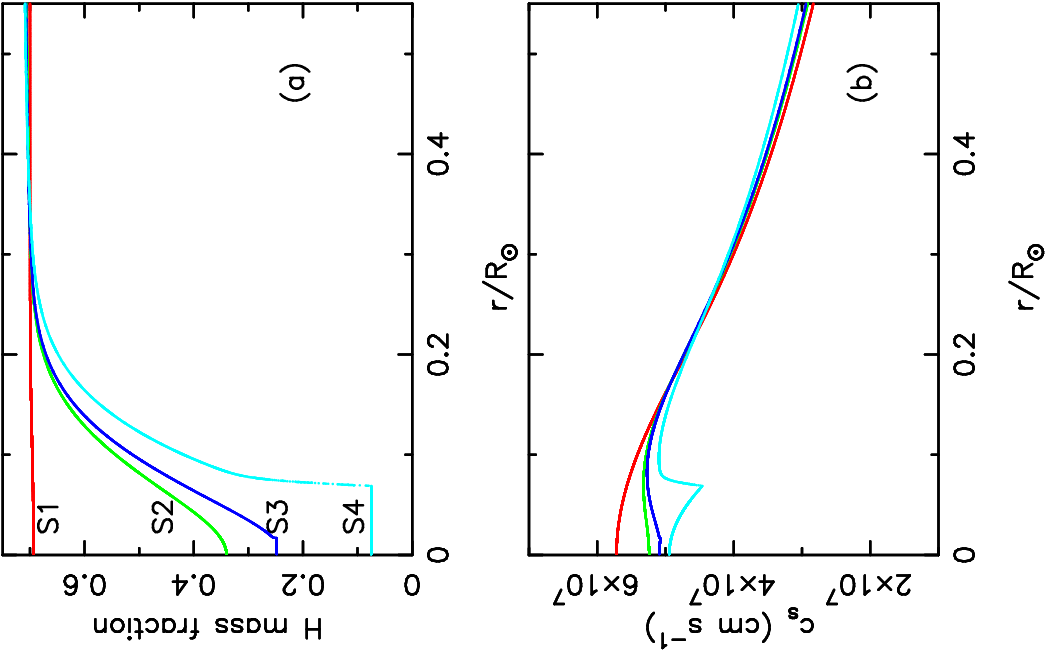}
\includegraphics[scale=0.5, angle=-90]{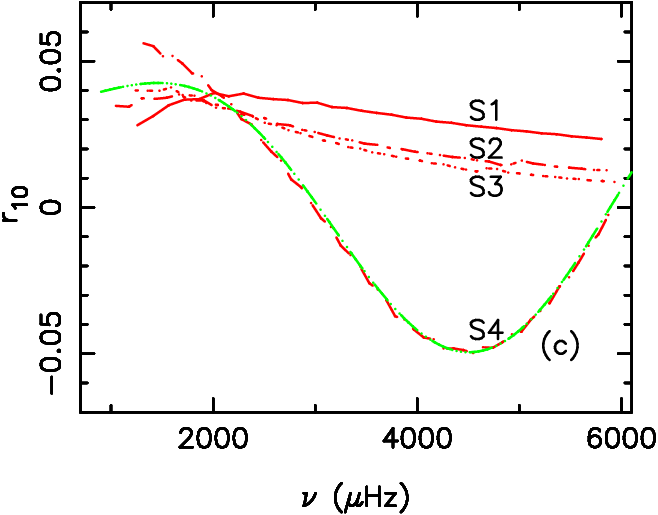}
\caption{\textbf{Panels (a) and (b) show radial distributions of H mass 
fraction and adiabatic sound speed of models with $M=1.09$ \dsm{},
respectively. The symbols S1, S2, S3, and S4 indicate
different evolutionary stages. Panel (c) represents distributions of
the ratio \drb{} of the models as a function of frequency. In panel (c),
the red lines correspond to \drb{} of models, while the green line 
shows the results of equation (\ref{eqb}).} } 
\label{fig3}
\end{figure}

\textbf{When} angular frequencies of modes are larger than
\textbf{a critical frequency $\omega_{0}$,} the modes can 
penetrates into the convective core \textbf{of stars}. 
Assuming that the effects of the convective core on oscillations
is related to $-A\cos(\omega_{0}t)$, where $A$ is a free parameter,
\cite{yang15b} obtained the \textbf{equation} (\ref{eqb}) as 
the result of Fourier transform of $-A\cos(\omega_{0}t)$. 
Thus the \textbf{equation} (\ref{eqb}) is invalid for stars
whose core is \textbf{radiative.} Figure \ref{fig2} shows 
the distributions of H mass fraction, adiabatic sound speed,
and \drb{} of \textbf{core-radiative models in different 
evolutionary stages.} The ratio \drb{} decrease with increase
in frequency. The distributions can not be reproduced by
\textbf{equation (\ref{eqb}). The core of model S2 in Figure
\ref{fig3} is also radiative. The distribution of \drb{} 
of the model cannot be reproduced by equation (\ref{eqb}) too. }

\begin{figure}
\centering
\includegraphics[scale=0.5, angle=-90]{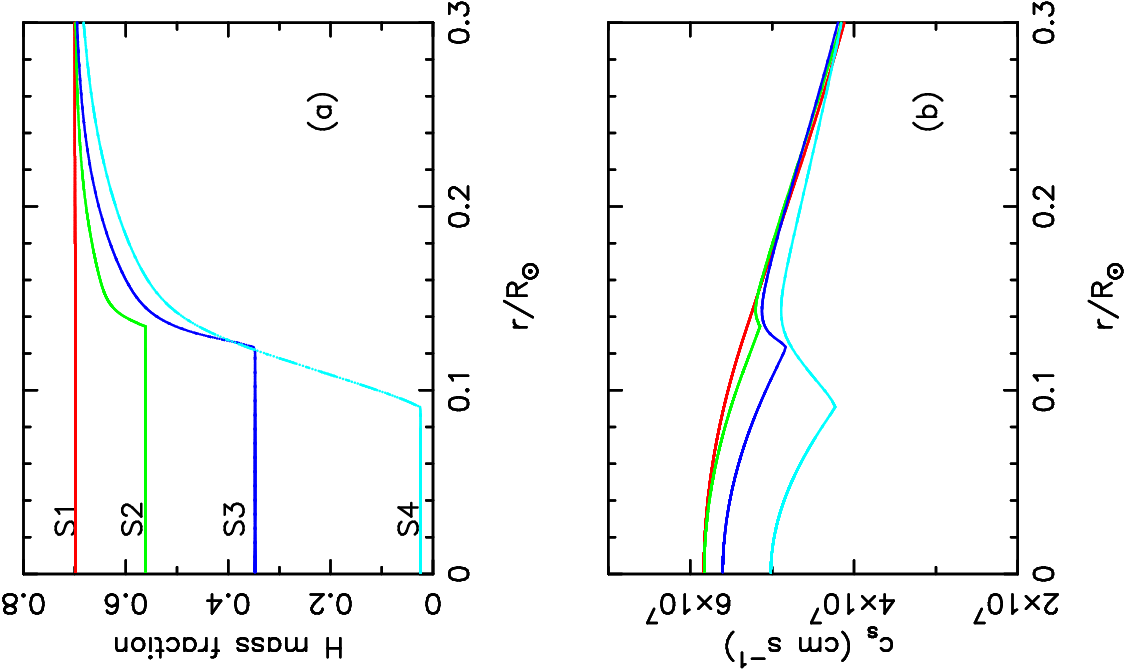}
\includegraphics[scale=0.5, angle=-90]{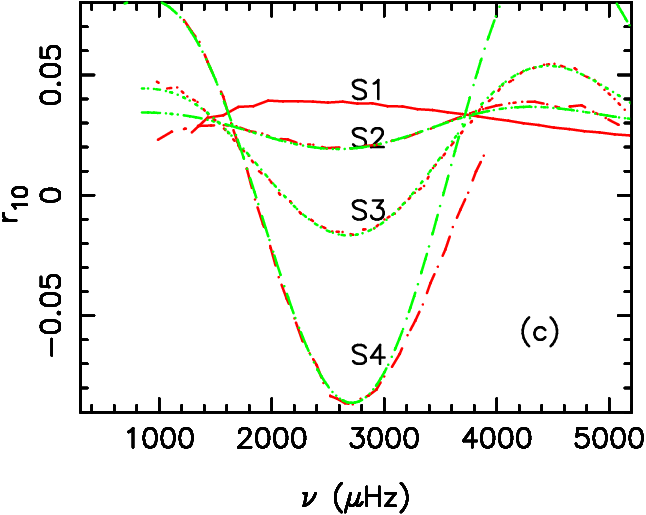}
\caption{\textbf{Similar to figure \ref{fig3}, but for the} 
models with $M=1.16$ \dsm{} and $\delta_{\rm ov}=0.2$. 
The distribution of \drb{} of model S3 is almost completely 
reproduced by \textbf{equation} (\ref{eqb}) with $\nu_{0}=2590$
\dhz{} which is approximately equal to $2600$ \dhz{} determined 
by equation (\ref{rc1}). \textbf{The cores of these models are convective. } } 
\label{fig4}
\end{figure}

\begin{figure}
\centering
\includegraphics[scale=0.5, angle=-90]{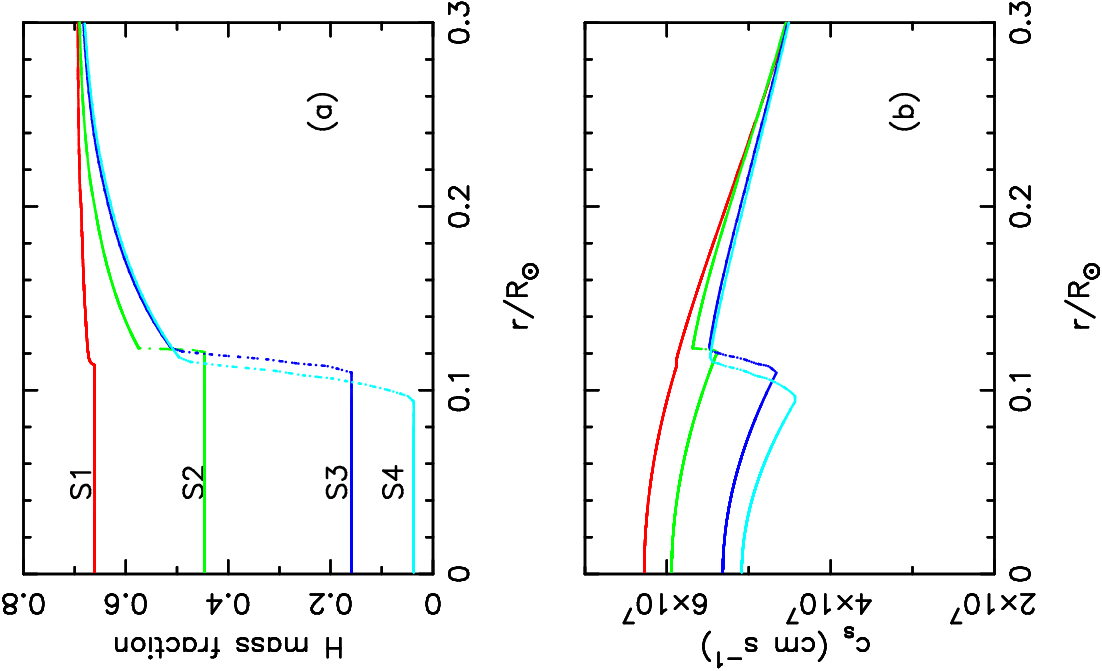}
\includegraphics[scale=0.5, angle=-90]{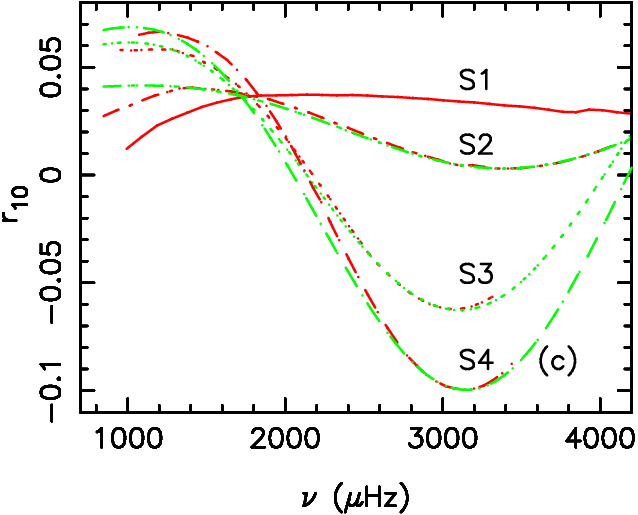}
\caption{Similar to figure \ref{fig3}, but for the models with $M=1.40$ \dsm{}
and $\delta_{\rm ov}=0$. \textbf{The cores of these models are convective. } }
\label{fig5}
\end{figure}

\cite{roxb04, roxb07} pointed out that \textbf{the discontinuity
in density} at the boundary of a convective core can distort acoustic
wave propagation in stellar interior, producing a reflected wave.
The effects of the convective core on oscillations are related
to the fact that the discontinuity reflects acoustic waves.
\textbf{Therefore,} the \textbf{equation} (\ref{eqb}) is invalid for stars
with a convective core but without the discontinuity in density 
or sound-speed at the edge of the convective core. \textbf{ 
Model S3 in Figure \ref{fig3} has a small convective core but  
has no an obvious discontinuity in density or sound speed at 
the edge of the convective core (see Figure \ref{fig3}). 
The distribution of \drb{} of the model cannot be reproduced
by equation (\ref{eqb}). While the model S4 has an obvious 
sound-speed discontinuity at the edge of the convective core.
The distribution of \drb{} of model S4 is almost reproduced by 
equation (\ref{eqb}). The cores of models S1 in Figures \ref{fig4}
and \ref{fig5} are also convective, but there is no an obvious 
discontinuity in density or sound speed of the models. 
The distributions of \drb{} of the models cannot be reproduced
by equation (\ref{eqb}) too. While the distributions of \drb{}
of models with a convective core and an obvious discontinuity
in sound speed at the edge of the convective core are almost
reproduced by equation (\ref{eqb}) (see Figures \ref{fig4} and \ref{fig5}). }

The modes with $l=0$ are considered to be able to reach 
the center of a star. According to equation (\ref{tp}), it is more
difficult to arrive at the convective core for the modes with
$l\geq2$ than for the modes with $l=1$. Thus the frequency
$\nu_{0}=\omega_{0}/2\pi$ could be the frequency of 
the $l=1$ mode whose inner turning point is located on the boundary
between the radiative region and the overshooting region of the 
convective core. \cite{roxb94, roxb99, roxb04, roxb07} show that 
there are periodic variations in small separations with period 
determined approximately by the acoustic diameter of the 
convective core, i.e., the period
\begin{equation}
T_{c} \approx 2\int_{0}^{r_{c}}\frac{dr}{c_{s}}.
\label{tc}
\end{equation}
The larger the value of $r_{c}$, the longer the $T_{c}$, i.e.,
the smaller the frequency $\nu_{c}=1/T_{c}$. According to
equation (\ref{tp}), the larger the value of $r_{c}$, 
the smaller the frequency $\nu_{0}$. Thus the frequency $\nu_{c}$ 
of Roxburgh could be related to the frequency $\nu_{0}$. 
The value of sound speed decreases from $5.60\times10^{7}$ cm s$^{-1}$
to $4.85\times10^{7}$ cm s$^{-1}$ in the convective core of model S3
of star with $M=1.16$ \dsm{} (see Figure \ref{fig4}). 
If the $c_{s}$ in equation (\ref{tc}) is replaced by $c_{s}(r_{c})$, 
the value of $\nu_{c}$ can be estimated to be about $0.5 c_{s}(r_{c})/r_{c}$. 
From equation (\ref{rc1}), one can obtain $\nu_{0} = 0.45 c_{s}(r_{c})/r_{c}$
which is very close to $\nu_{c}$. 

In stellar interior, sound speed decreases with the increase
in radius. The value of $c_{s}(r_{c})$ varies with
the mass and the age of stars and is affected by overshooting.
The value of $c_{s}(r_{c})$ of the most of MS stars with 
masses between about 1.1 and 1.5 \dsm{} is mainly in the range
of $(4-6)\times10^{7}$ cm s$^{-1}$ (see Figures \ref{fig4}
and \ref{fig5}). \textbf{For example, for}
a star with $M=1.16$ \dsm{}, $X_{i}=0.7$, 
$Z_{i}=0.02$, and $\delta_{\rm ov}=0.2$, when it evolves from 
central hydrogen abundance $X_{c}=0.59$ to $X_{c}=0.16$, 
the value of $c_{s}(r_{c})$ decreases from about $5.2\times10^{7}$
cm s$^{-1}$ to $4.5\times10^{7}$ cm s$^{-1}$; \textbf{for} a star 
with $M=1.4$ \dsm{}, $X_{i}=0.7$, $Z_{i}=0.02$, and $\delta_{\rm ov}=0$,
when it evolves from $X_{c}=0.5$ to $X_{c}=0.1$, the value
of $c_{s}(r_{c})$ decreases from about $5.5\times10^{7}$ cm s$^{-1}$
to $4.5\times10^{7}$ cm s$^{-1}$. Therefore, in the most of 
MS stage of stars with masses between about 1.1 and 1.5 \dsm{}, 
\textbf{taken a model in the middle stage of MS as a reference,}
one can assume that there is a change of $10\%$ in $c_{s}(r_{c})$, 
i.e., $c_{s}(r_{c})\sim(5.0\pm0.5)\times10^{7}$ cm s$^{-1}$.

\textbf{For our sample, the relative uncertainty of $\nu_{0}$ 
determined by chi-square fitting from observed data is between
about $2.4\%$ and $6.2\%$. But our sample is small. The relative 
uncertainty of $\nu_{0}$ of other stars might be larger than 
$6.2\%$. In order to estimate the uncertainty of the estimated 
$r_{c}$ of other stars when this method is applied to the stars, 
we assume that the relative uncertainty of $\nu_{0}$ for 
other stars is of the order of $10\%$ and apply the uncertainty
of $10\%$ to all cases.} As a consequence, the \textbf{relative} 
uncertainty of the estimated $r_{c}$ is about $20\%$. 
Table \ref{tab3} shows that the values of the radius of 
the convective core determined by equations (\ref{eqb}) 
and (\ref{rc2}) from observed frequencies of different stars
are in good agreement with those obtained from the best models 
of the stars. 

\subsection{Summary}
Combining equation (\ref{eqb}), we propose here for the first time 
using formula (\ref{rc2}) to estimate the radius of the convective
core including overshooting region of MS stars with masses
between about 1.1 and 1.5 \dsm{} from 
observed frequencies and ratios. The estimated values of
the radius of the convective core of four stars are 
consistent with those of the best models of the four stars. 
Using the observed frequencies and ratios of KIC 9812850,
equations (\ref{eqb}) and (\ref{rc2}) predict the radius
of the convective core of KIC 9812850 is $\mathbf{0.140\pm0.028}$ \dsr{}.
In order to confirm this prediction, we constructed 
a grid of evolutionary tracks. Basing on finding
the maximum likelihood of models, we obtained 
the best-fit model of KIC 9812850 with $M=1.48$ \dsm{}, 
$R=1.867$ \dsr{}, \teff{}$=6408$ K, $L=5.28$ \dsl{}, $t=2.606$ Gyr,
$r_{c}=0.149$ \dsr{}, and \dov{}$=0.2$. The best model can 
reproduce asteroseismic and non-asteroseismic characteristics 
of KIC 9812850. The value of the radius of the convective core 
of the best-fit model is in good agreement with that predicted
by formula (\ref{rc2}). Equations (\ref{eqb}) and (\ref{rc2})
aid in understanding the structure of stars directly from the 
observed frequencies.

\acknowledgments
The author thanks the anonymous referee for helpful
comments that helped the author improve this work, as well
as the support from the NSFC 11273012, 11273007,
the Fundamental Research Funds for the Central Universities,
and the HSCC of Beijing Normal University.

\clearpage

\clearpage

\begin{table}
\begin{center}
\caption[]{The range of the input parameters for the evolutionary tracks.}
\label{tab1}
\begin{tabular}{ccccc}
  \hline\noalign{\smallskip}
  \hline\noalign{\smallskip}
   Variable         & Minimum  & Maximum & Resolution  \\
   \hline\noalign{\smallskip}
     $M$/M$_\odot$  & 1.00     & 1.60  & $\leq$0.02    \\
      $\alpha$      & 1.65     & 2.05  &  0.1     \\
      $\delta_{\rm ov}$ &  0.0 & 1.8   &  0.2     \\
      $Z_{i}$       & 0.010    & 0.040 & 0.002   \\
   \hline\noalign{\smallskip}

  \noalign{\smallskip}
\end{tabular}
\end{center}
\end{table}

\clearpage

\begin{table}
\begin{center}
\renewcommand\arraystretch{1.0}
\caption[]{Parameters of models for KIC 9812850. }
\label{tab2}
\begin{tabular}{p{0.70cm}cccccccccccccc}
  \hline\hline\noalign{\smallskip}
   Model & $M$ & \teff{}& $ L$ & $R$& Age & $Z_{i}$ & $X_{i}$ & $\alpha$ & \dov{}
   & $r_{\mathrm{c}}$  & $\nu_{0}^{m}$ & \dcn{} & \dcc{} & $\mathcal{L}$ \\

  & (\dsm{}) & (K) & (\dsl{})  & (\dsr{}) & (Gyr) & &  & & & (\dsr{})& ($\mu$Hz) & & \\
  \hline\hline\noalign{\smallskip}
 M1 & 1.44 & 6267 & 4.75 & 1.851 & 2.861 & 0.024 & 0.728 & 1.75 & 0.2 & 0.144 &  2364 &  1.5 & 0.7 & 1.8$\times10^{-22}$ \\
 M2 & 1.46 & 6188 & 4.58 & 1.864 & 2.950 & 0.028 & 0.724 & 1.75 & 0.2 & 0.147 &  2248 &  1.0  &  2.0& 9.1$\times10^{-19}$ \\
 M3 & 1.48 & 6408 & 5.28 & 1.867 & 2.606 & 0.024 & 0.728 & 1.95 & 0.2 & 0.149 &  2246 &  0.7  &  1.6& 1.0$\times10^{-15}$ \\
 M4 & 1.50 & 6320 & 5.07 & 1.880 & 2.620 & 0.030 & 0.722 & 1.95 & 0.2 & 0.152 &  2184 &  1.0  &  1.8& 1.6$\times10^{-18}$ \\
 \noalign{\smallskip}\hline\hline
\end{tabular}
\end{center}
{Note. The symbol $r_{\mathrm{c}}$ represents the radius of convective core of models; 
while the $\nu_{0}^{m}$ indicates the frequency at which ratios \dra{} and \drb{} of models 
reach the minimum.}
\end{table}

\begin{table}
\begin{center}
\caption[]{The radius of the convective core of five stars.}
\centering
\label{tab3}
\begin{tabular}{cccc}
  \hline\hline\noalign{\smallskip}
     Star    & $\nu_{0}$/$\mu$Hz & $r_{c}$/$R_{\odot}$ & $r_{c,model}$/$R_{\odot}$ \\  
  \hline\noalign{\smallskip}
 KIC 6225718 & $5764\pm312$      & $0.056\pm0.011^{(a)}$& 0.056  \\
 KIC 9812850 & $2309\pm141$      & $0.140\pm0.028$     & $0.149$  \\ 
   HD 49933  & $1920\pm46$       & $0.168\pm0.034$     & 0.170  \\
 KIC 2837475 & $913\pm36$        & $0.354\pm0.070$     & 0.358  \\
 KIC 11081729 &$795\pm21$        & $0.406\pm0.081$     & 0.370  \\
 \noalign{\smallskip}\hline\hline
\end{tabular}
\end{center}
$^{(a)}${The uncertainty is estimated by assuming that there is an uncertainty
of $10\%$ in $c_{s}$ and $\nu_{0}$. The values of $r_{c}$ are estimated by using
equation (\ref{rc2}), while the values of $r_{c,model}$ are obtained from 
the best models of the stars.}
\end{table}


\begin{thebibliography}{}

\bibitem[Alexander \& Ferguson(1994)]{alex94} Alexander, D. R., \& Ferguson, J. W. 1994, ApJ, 437, 879
\bibitem[Ammons et al.(2006)]{ammo06} Ammons, S. M., Robinson, S. E.,
Strader, J., Laughlin, G., Fischer, D., \& Wolf, A. 2006, ApJ, 638, 1004
\bibitem[Appourchaux et al.(2012)]{appo12}
Appourchaux, T., Chaplin, W. J., Garc\'{i}a, R. A., et al. 2012, A\&A, 543, A54
\bibitem[Basu et al.(2010)]{basu10} Basu, S., Chaplin, W. J., \& Elsworth, Y. 2010, ApJ, 710, 1596
\bibitem[Benomar et al.(2009)]{beno09} Benomar, O., Baudin, F., Campante, T. L. et al. 2009, A\&A, 507, L13
\bibitem[Brand$\tilde{a}$o et al.(2014)]{bran14} Brand$\tilde{a}$o, I. M., Cunha, M. S., 
\& Christensen-Dalsgaard, J. 2014, MNRAS, 438, 1751
\bibitem[Brand$\tilde{a}$o et al.(2010)]{bran10}
Brand$\tilde{a}$o, I. M., Cunha, M. S., Creevey, O. L.,
\& Christensen-Dalsgaard, J. 2010, AN, 331, 940
\bibitem[Bruntt et al.(2012)]{brun12}
Bruntt, H., Basu, S., Smalley, B., et al. 2012, MNRAS, 423, 122
\bibitem[Chaplin et al.(2014)]{chap14}
Chaplin, W. J., Basu, S., Huber, D., et al. 2014, ApJS, 210, 1
\bibitem[Christensen-Dalsgaard \& Houdek(2010)]{chri10}
Christensen-Dalsgaard, J., \& Houdek, G. 2010, Ap\&SS, 328, 51
\bibitem[Cunha \& Metcalfe(2007)]{cunh07} Cunha, M. S., \& Metcalfe, T. S. 2007, ApJ, 666, 413
\bibitem[Cunha \& Brand$\tilde{a}$o(2011)]{cunh11} 
Cunha, M. S., \& Brand$\tilde{a}$o, I. M. 2011, \aap, 529, A10
\bibitem[Deheuvels et al.(2010)]{dehe10}
Deheuvels, S., Bruntt, H., Michel, E., et al. 2010, \aap, 515, A87
\bibitem[Demarque et al.(1994)]{dema94}
Demarque, P., Sarajedini, A., Guo, X.-J. 1994, ApJ, 426, 165
\bibitem[Droege et al.(2006)]{droe06}
Droege, T. F., Richmond, M. W., Sallman, M. P., Creager, R. P. 2006, PASP, 188, 1666
\bibitem[Flower(1996)]{flow96} Flower, P. J. 1996, ApJ, 469, 355

\bibitem[Ge et al.(2014)]{ge14} Ge, Z. S., Bi, S. L., Li, T. D.,
Liu, K., Tian, Z. J., Yang, W. M., Liu, Z. E., Yu, J. 2014, MNRAS, 447, 680
\bibitem[Grevesse \& Sauval(1998)]{grev98}
Grevesse, N., \& Sauval, A. J. 1998, in Solar Composition and Its Evolution,
ed. C. Fr$\ddot{o}$hlich et al. (Dordrecht: Kluwer), 161

\bibitem[Guenther(1994)]{guen94} Guenther, D. B. 1994, ApJ, 422, 400
\bibitem[Guenther et al.(2014)]{guen14}
Guenther, D. B., Demarque, P., \& Gruberbauer, M. 2014, ApJ, 787, 164
\bibitem[Iglesias \& Rogers(1996)]{igle96} Iglesias, C., Rogers, F. J. 1996, ApJ, 464, 943
\bibitem[Kharchenko(2001)]{khar01} Kharchenko, N. V. 2001, KFNT, 17, 409
\bibitem[Kjeldsen et al.(2008)]{kjel08}
Kjeldsen, H., Bedding, T. R., \& Christensen-Dalsgaard, J. 2008, ApJL, 683, L175
\bibitem[Liu et al.(2014)]{liu14} Liu, Z., Yang, W., Bi, S., et al. 2014, ApJ, 780, 152
\bibitem[Mazumdar et al.(2014)]{mazu14} Mazumdar, A., Monteiro, M. J. P. F. G., Ballot, 
J., et al. 2014, ApJ, 782, 18
\bibitem[Metcalfe et al.(2014)]{metc14} Metcalfe, T. S., et al. 2014, ApJS, 214, 27
\bibitem[Molenda-Zakowicz et al.(2013)]{mol13}
Molenda-Zakowicz, J., Sousa, S. G., Frasca, A., et al. 2013, MNRAS, 434, 1422
\bibitem[Pinsonneault et al.(1989)]{pins89}
Pinsonneault, M. H., Kawaler, S. D., Sofia, S., \& Demarqure, P. 1989, ApJ, 338, 424
\bibitem[Rogers \& Nayfonov(2002)]{roge02} Rogers, F. J., \& Nayfonov, A. 2002, ApJ, 576, 1064

\bibitem[Roxburgh \& Vorontsov(1994)]{roxb94} Roxburgh, I. W., \& Vorontsov, S. V. 1994, MNRAS, 267, 297
\bibitem[Roxburgh \& Vorontsov(1999)]{roxb99} Roxburgh, I. W., \& Vorontsov, S. V. 1999, 
in Stellar Structure: Theory and Test of Connective Energy Transport, ed. A. Gim\'{e}nez, E. F. Guinan, 
\& B. Montesinos, ASP Conf. Ser. (San Francisco), 173, 257
\bibitem[Roxburgh \& Vorontsov(2000a)]{roxb00a} Roxburgh, I. W., \& Vorontsov, S. V. 2000a, MNRAS, 317, 141
\bibitem[Roxburgh \& Vorontsov(2000b)]{roxb00b} Roxburgh, I. W., \& Vorontsov, S. V. 2000b, MNRAS, 317, 151
\bibitem[Roxburgh \& Vorontsov(2001)]{roxb01} Roxburgh, I. W., \& Vorontsov, S. V. 2001, MNRAS, 322, 85
\bibitem[Roxburgh \& Vorontsov(2003)]{roxb03} Roxburgh, I. W., \& Vorontsov, S. V. 2003, A\&A, 411, 215
\bibitem[Roxburgh \& Vorontsov(2004)]{roxb04} Roxburgh, I. W., \& Vorontsov, S. V. 2004, 
in ESA Special Publication, Vol. 538, Stellar Structure and Habitable Planet Finding, 
ed. F. Favata, S. Aigrain, \& A. Wilson (Noordwijk: ESA), 403
\bibitem[Roxburgh \& Vorontsov(2007)]{roxb07} Roxburgh, I. W., \& Vorontsov, S. V. 2007, MNRAS, 379, 801

\bibitem[Silva Aguirre et al.(2011)]{silv11}Silva Aguirre, V., Ballot, J., Serenelli, A. M., \& Weiss, A. 2011, A\&A, 529, A63
\bibitem[Silva Aguirre et al.(2013)]{silv13}Silva Aguirre, V., Basu, S., Brand$\tilde{a}$o, I. M., et al. 2013, ApJ, 769, 141
\bibitem[Spergel et al.(2007)]{sper07} Spergel, D. N., Bean, R., Dore, O., et al., 2007, ApJS, 170, 377

\bibitem[Thoul et al.(1994)]{thou94} Thoul, A. A., Bahcall, J. N., Loeb, A. 1994, ApJ, 421, 828
\bibitem[Tian et al.(2014)]{tian14}Tian, Z. J., Bi, S. L., Yang, W. M., et al. 2014, MNRAS, 445, 2999
\bibitem[Wright et al.(2003)]{wright03} Wright, C. O., Egan, M. P., Kraemer, K. E., Price, S. D. 2003, AJ, 125, 359
\bibitem[Yang(2015)]{yang15b} Yang, W. 2015, arXiv:1508.00955
\bibitem[Yang(2016)]{yang16} Yang, W. 2016, ApJ, 821, 108
\bibitem[Yang \& Bi(2007a)]{yang07a} Yang, W., \& Bi, S. 2007a, ApJL, 658, L67
\bibitem[Yang \& Bi(2007b)]{yang07b} Yang, W., \& Bi, S. 2007b, \aap, 472, 571
\bibitem[Yang \& Meng(2010)]{yang10a} Yang, W., \& Meng, X. 2010, NewA, 15, 367
\bibitem[Yang et al.(2012)]{yang12} Yang, W., Meng, X., Bi, S., Tian, Z., Liu, K.,
Li, T., Li, Z. 2012, MNRAS, 422, 1552
\bibitem[Yang et al.(2015)]{yang15} Yang, W., Tian, Z., Bi, S., Ge, Z., Wu, Y., \& Zhang, J. 2015, MNRAS, 453, 2094

\end{thebibliography}
\end{document}